\documentclass[twocolumn]{aa}
\usepackage{graphicx}
\usepackage{ifthen}
\usepackage{natbib}
\usepackage{txfonts}

\def\CompactFigs{0}

\def\ltsima{$\; \buildrel < \over \sim \;$}
\def\simlt{\lower.5ex\hbox{\ltsima}}
\def\gtsima{$\; \buildrel > \over \sim \;$}
\def\simgt{\lower.5ex\hbox{\gtsima}}
\def\kpc{{\rm\,kpc}}

\begin{document}
   \title{Tracing the Sgr Stream with 2MASS}

   \subtitle{Detection of Stream stars around Outer Halo Globular Clusters
   \thanks{This publication makes use of data products from the Two Micron All 
   Sky Survey, which is a joint project of the University of Massachusetts and 
   the Infrared Processing and Analysis Center/California Institute of 
   Technology, funded by the National Aeronautics and Space Administration and 
   the National Science Foundation}}

   \author{M. Bellazzini\inst{1}, R. Ibata\inst{2}, F.R. Ferraro\inst{3}, V.
   Testa\inst{4}}

   \offprints{M. Bellazzini}

\institute{INAF - Osservatorio Astronomico di Bologna, Via Ranzani 1, 40127, 
          Bologna, Italy                
              \email{bellazzini@bo.astro.it}
	  \and  
	   Observatoire de Strasbourg, 67000 Strasbourg, France
         \email{ibata@astro.u-strasbg.fr}
	 \and 
	 Dipartimento di Astronomia, Universit\`a di Bologna, 
	 Via Ranzani 1, 40127, Bologna, Italy
	 \email{ferraro@bo.astro.it}
	 \and 
         INAF - Osservatorio Astronomico di Roma, Via Frascati 33, 00040, 
          Monte Porzio Catone, Italy                
              \email{testa@mporzio.astro.it}	      }

   \date{Received February 25, 2003; Accepted April 24, 2003 }

\abstract{We use  infrared Color Magnitude Diagrams  from the 2-Micron
All-Sky  Survey (2MASS)  to search  for stars  belonging to  the tidal
stream of  the Sagittarius dwarf  spheroidal galaxy (Sgr  dSph) around
selected   Galactic  globular   clusters.   Statistically  significant
detections  are  presented  for  the  cases of  Pal~12  and  NGC~4147,
strongly supporting  the idea that these clusters  are associated with
the Sgr  Stream and that  they were previous  members of the  Sgr dSph
galaxy.    }

   \maketitle
%

\section{Introduction}

There is  clear evidence that the Sagittarius  dwarf Spheroidal galaxy
(Sgr  dSph; \citealt{s1,s2})  has lost  (and is  currently  loosing) a
significant fraction  of its stars under  the strain of  the Milky Way
tidal field.   The lost stars  seem able to remain  coherently aligned
with the  orbital path of the Sgr  dSph as a long-lived  ($>$ few Gyr;
see \cite{katy2,orb}) stellar relic, known as the Sgr Stream. Parts of
the Sgr  Stream have been detected  by many different  groups and with
many different techniques, up to  $> 100\degr$ away from the main body
of  the galaxy  \citep{sgrm55,  mom98, maj99,  ami1, carb,  orb-2mass,
kundu, david,  vivas, yanni, ivez}.  Direct  identifications, based on
the  similarity of  the CMD  in the  main body  of the  galaxy  and in
portions of  the Stream  $\sim 160 \degr$  away in  galactic longitude
have been provided by \cite{sdss}.

\citet[][hereafter  IL98]{orb}  and  \cite{carb}  have  simulated  the
evolution of  the Sgr  dSph over several  orbital periods (P  $\sim 1$
Gyr), computing  the orbit  of the galaxy  as well as  the phase-space
distribution  of  the debris  under  different  assumptions about  the
flattening of the CDM halo.  The initial conditions of the simulations
were based on  the known position and radial velocity  of Sgr dSph and
on its proper motion as  estimated by \cite{s2,carb}.  The orbit has a
planar  rosette structure,  with  the  pole of  the  orbit located  at
[$\ell=90^\circ$,  $b=-13^\circ$]  (i.e. a  nearly  polar orbit),  and
peri-   and   apo-Galactic   distances   of  $15\kpc$   and   $60\kpc$
respectively.  The derived orbit has successfully been used to predict
the  observed position  (\citealt{carb,orb-sds})  and radial  velocity
(\citealt{yanni2}) of the Sgr Stream.

In a previous paper  (\citealt{pap1}, hereafter BFI03) we compared the
IL98 orbit  during the last orbital  period with a  subset of galactic
globular clusters,  namely those having $10\kpc  \le R_{GC}\le 40\kpc$
which  we termed  Outer Halo  (OH)  globulars, in  the phase-space  of
Galactocentric cartesian coordinates and radial velocity ($X,Y,Z,V_r$:
see  BFI03  for  further  details).   We  found  that  the  region  of
phase-space occupied by the  Sgr Stream hosts a significant overdensity
of OH  clusters, thus  some of them  have to be  physically associated
with the  Stream and were former  members of the Sgr  dSph.  While the
association of at least 4 OH  clusters with the Stream is quite firmly
established, the result is statistical in nature and we cannot confirm
the  actual  membership  of   individual  clusters  with  the  adopted
technique.   Hence, we  simply  provided a  list  of candidate  Stream
members ranked  according to their proximity  to the Sgr  orbit in the
considered  phase-space.  Strong  support for  individual associations
may be  provided by a  direct detection of  Sgr Stream stars  around a
candidate  cluster  and located  at  the  same  distance from  us,  as
provided,  for   instance,  by  \cite{dav12}   in  the  case   of  our
first-ranked candidate Pal~12 (see also \citealt{pal12}).

This kind of search requires the  sampling of wide areas of sky, since
the Stream is a \emph{very}  low surface brightness structure, and the
use of suitable  tracer stars that can be singled  out from the strong
contamination   provided  by  ordinary   Galactic  field   stars  (see
\citealt{carb,yanni,ivez,sdss,vivas,orb-2mass}   for   the   different
tracers  used to  date).  Here  we use  stars in  the upper  Red Giant
Branch (RGB) from the  near infrared Color-Magnitude Diagrams (CMD) of
wide areas of  sky provided by the 2-Micron  All-Sky Survey (2MASS) to
search  for Stream stars  around the  top-ranked candidates  listed in
Tab.~1  of BFI03.   Sgr Stream  stars  at the  ``right'' distance  are
detected around  Pal~12, thus  confirming the result  by \cite{dav12},
and around NGC~4147, a completely new detection. The present study has
to be considered as a preliminary ``pilot'' analysis.  We will perform
a  deeper  and more  complete  search  once  the final  all-sky  2MASS
catalogue is released.

The plan of the paper is the following: in \S2 we present the data and
the method;  in \S3 and  in \S4 we  describe the results  obtained for
Pal~12 and  NGC~4147, respectively;  in \S5 we  deal briefly  with the
case of the other considered candidates; \S6 gives a short summary and
a discussion of the reported results.


\section{Data and method}

All the photometric data are from the Point Source Catalogue (PSC) of the Second
Incremental Data Release of 2MASS\footnote{see Cutri, R.M., et al., 2000, 
Explanatory Supplement to the 2MASS Second Incremental Data Release, {
http://www.ipac.caltech.edu/2mass/releases/second/doc/expsup.html}}
and consist  of calibrated J,H,K$_S$\footnote{From now on  we drop the
$S$  index from  K$_S$,  for brevity.}   magnitudes  and positions  of
sources classified  as stellar and without  any detected contamination
from other  nearby sources. Magnitudes  and colors of all  the retrieved
sources   have   been   corrected   for  interstellar   extinction   by
interpolating  in the  IRAS-DIRBE  reddening maps  of \cite{iras}  and
adopting the extinction laws  by \cite{sm79}.  The present sky coverage
of 2MASS is  quite patchy, hence there are regions  of sky that cannot
be scrutinized. We typically  select areas of $20\degr \times 20\degr$
toward interesting directions, but in all cases large unobserved zones
are included  in the selected  areas. Typically, the  effective sampled
area  ranges  from $\sim  \frac{1}{4}$  to  $\sim  \frac{2}{3}$ of  the
selected area (see, for example, Fig.~4 and Fig.~5, below).  Distances
and  other  parameters  for  globular  clusters  are  taken  from  the
compilation by \cite{harris}.

   \begin{figure}
   \centering
   \ifthenelse{\CompactFigs=0}
   {\includegraphics[width=9cm]{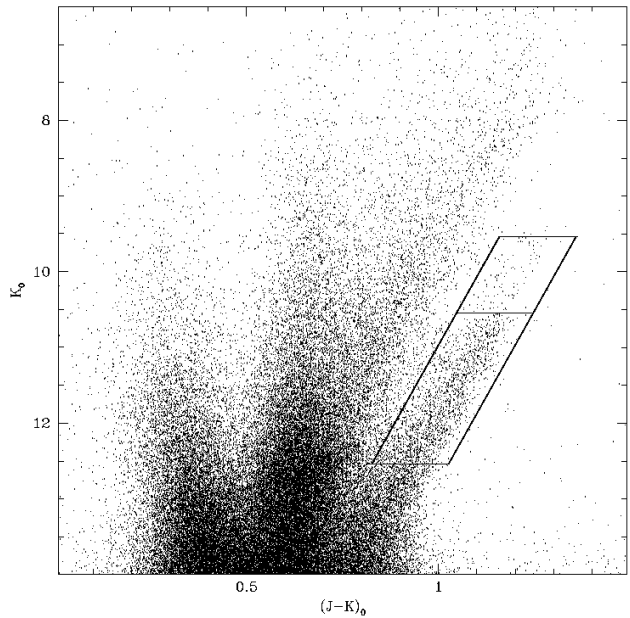}}
   {\includegraphics[width=9cm]{fig1.ps}}{}
   \caption{$K_S$,$J-K_S$ CMD of a $\sim 4\degr \times 4\degr$ area around the
   center of the Sgr dSph galaxy. The upper RGB of Sgr stands out very clearly
   at $J-K_S>1$, to the red of all the features formed by galactic field stars.
   The lower box encloses the Sgr RGB stars, the upper box encloses Sgr bright
   AGB stars.}
    \end{figure}
%

Fig.~1 displays  the K,J-K  CMD of  a $4 \degr  \times 4  \degr$ field
centered on the  point of maximum density of the main  body of the Sgr
dSph.  A similar  field  has been  used  by \cite{cole}  to study  the
metallicity of the galaxy and  virtually the {\em same} field has been
considered  by  \cite{alard}  in   his  analysis  of  the  metallicity
gradient. As  noted by these authors  the brighter part of  the RGB of
the Sgr  dSph stands out very clearly  in this diagram, to  the red of
any feature  due to galactic  field stars in the  region $(J-K)_0>0.8$
and $K_0\simlt 12$ (see \cite{cole} and \cite{alard} for discussion of
the field  features in the  CMD). The lower  part of the  reported box
encloses  the more  clearly visible  part of  the Sgr  RGB.  The upper
(smaller) sector  of the box  encloses bright Asymptotic  Giant Branch
(AGB) stars of Sgr.  We will use  only RGB stars as tracers of the Sgr
stellar population but we maintain also  the upper sector of the box to
check also if  rarer AGB stars fall in our  selection box. The reddest
Sgr stars from 2MASS have  been already used by \cite{orb-2mass} for a
pole-count analysis that detected an  overdensity of such stars in the
plane of the sky in the directions predicted by the IL98 orbit.

The detection  method is quite simple:  since Sgr RGB  stars stand out
clearly against  galactic field stars  in the considered CMD,  we will
look if similar stars can be  found also in wide fields surrounding OH
globulars that  are candidate stream  members according to  BFI03. The
RGB  feature,  if present,  has  to be  shifted  in  magnitude by  the
difference  in true  distance  moduli  between the  Sgr  dSph and  the
considered cluster  ($\Delta\mu_0$). Therefore, we  will simply report
the Selection  Box (SB) shown  in Fig.~1 in  the CMDs of  the selected
area  and  of  a suitable  Control  Field  (CF)  applying a  shift  by
$\Delta\mu_0$.   Then  we  will   check  if  there  is  a  significant
overdensity of  stars in the SB  of the selected area  with respect to
the CF.  It  is important to note that the SB  has been constructed to
include  also the  bluest stars  in the  Sgr RGB,  in order  to remain
sensitive to  the searched feature  notwithstanding the effect  of the
metallicity/population radial gradient observed in Sgr \citep{alard}.

Since we are looking for very low-surface-brightness structures, it is
of   the  utmost   importance  to   avoid  any   possible   source  of
contamination. To  this end, within the  SB we will  select only stars
having  $(J-K)_0>0.95$, to avoid  the inclusion  of the  reddest stars
from the local M dwarf plume  at $(J-K)_0\sim 0.8$, and we will reject
stars with $K_0>12$,  since in this range of  magnitude a sparse cloud
of red  sources (possibly misclassified  galaxies\footnote{The feature
is more evident  in less crowded fields with respect  to that shown in
Fig.~1;  see Fig.~2 and  Fig.~3.}) provides  additional contamination.
For this  reason, the adopted  technique is nearly insensitive  to the
feature of interest for  $\Delta\mu_0\simgt +0.5$, i.e. the Sgr Stream
RGB can be  detected only up to  a distance of $\sim 32$  Kpc from us.
In  some cases  (e.g., NGC~4147)  the considered  clusters fall  in an
unobserved zone of the selected  area, thus contamination of the SB by
cluster stars  cannot occur. In  all other considered cases  we checked
the CMD of  the stars in the immediate vicinity of  the cluster (up to
$1 \degr$ from  its center) and we never found the  case of a possible
cluster star  falling in  the SB.  This  is due  to the fact  that the
considered clusters  are significantly more  metal poor than  the main
population of Sgr (see \citealt{cole,lorenzo} and references therein),
hence their  RGBs are  much bluer and  fall within the  Galactic field
features of the CMD.

In  this first attempt  we considered  the first  six clusters  of the
ranked list of candidates of  BFI03, these are, in order of increasing
phase-space  distance from  the IL98  orbit: Pal~12,  NGC~4147, Pal~5,
NGC~5634,  NGC~5053 and Ter~3.   Searching the  PSC catalogue  we have
found suitable selected areas for Pal~12 and NGC~4147. The present sky
coverage of 2MASS around  NGC~5634, Pal~5 and NGC~5053 is insufficient
to obtain  an effective sampled  area useful for the  present purpose.
Finally, we recently realized (Marsakov, private communication) that
the 2003 version of the \citet{harris} catalogue reports
significantly revised estimates of the reddening and distance modulus of 
Ter~3 [E(B-V)=0.72; $(m-M)_V=16.61$] with respect to the 1996 version that
was used by BFI03 [E(B-V)=0.32; $(m-M)_V=18.00$]. The new distance estimate
places the cluster quite far from the Sgr orbit. As a consequence, Ter~3
was removed from the list of candidates and it is not considered in the
present analysis. Thus, the test remains to be done
for  Pal~5, NGC~5634 and  NGC~5053 (these  cases are  shortly
discussed in \S5, below), while  in the following two sections we show
the results we have obtained for the cases of Pal~12 and NGC~4147.

   \begin{figure*}
   \centering
   \ifthenelse{\CompactFigs=0}
   {\includegraphics[width=18cm]{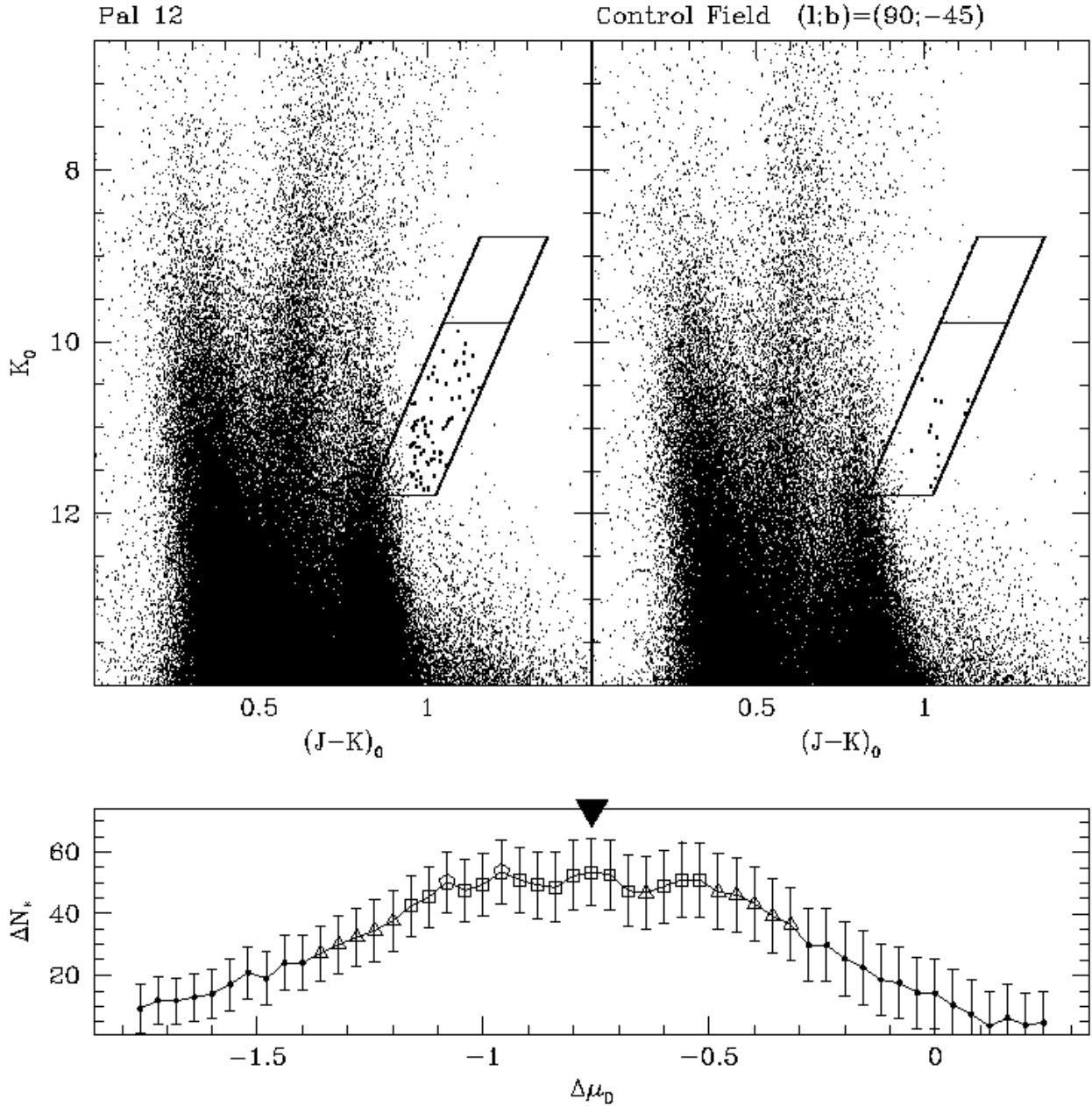}}
   {\includegraphics[width=18cm]{fig2.ps}}{}
   \caption{Upper panels: the CMD of the selected area around the position of
   Pal~12 (left upper panel) and of a Control Field located at similar Galactic
   latitude. The reported Selection Box has been shifted by 
   $\Delta\mu_0 = -0.76$ with respect to Fig.~1, to displace it to the distance of
   Pal~12. The selected stars are marked with heavier points. Lower panel:
   The normalized difference in the number of stars falling in the Selection Box
   as a function of $\Delta\mu_0$. Different symbols correspond to different
   values of $\Delta N_{\star}$ in units of $\sigma$ (filled points: $\Delta
   N_{\star}/\sigma <3$ , triangles: $3\le N_{\star}/\sigma <4$, squares:
   $4\le N_{\star}/\sigma <5$, etc.). The large filled triangle on the top of
   the panel marks the assumed $\Delta\mu_0$ for Pal~12.}
    \end{figure*}
%

\section{Pal 12}

To search  for Sgr  Stream stars around  this cluster,  whose galactic
coordinates  are $(l;b)=(30.512\degr;-47.681\degr)$,  we  selected the
area  comprised   in  the  ranges  $20\degr\le  l   \le  40\degr$  and
$-57\degr\le b  \le -37\degr$.  The effectively sampled  field is 71\%
of the selected  area.  In the upper panels of Fig.~2  the CMDs of the
Pal~12 field is compared to the CMD of a suitable nearby Control Field
at  similar  galactic  latitude   ($80\degr\le  l  \le  100\degr$  and
$-55\degr\le  b  \le  -35\degr$).   The difference  in  true  distance
modulus  between Pal~12  and the  core of  the Sgr  galaxy  is $\Delta
\mu_0=\mu_0(P12)-\mu_0(M54)=-0.76\pm  0.15$.  Hence  the  SB shown  in
Fig.~2 has been shifted by $\Delta \mu_0=-0.76$ in K to search for the
RGB feature at the distance of Pal~12.

Even  from a  visual inspection  it is  clear that  Sgr RGB  stars are
present in  the field surrounding the  cluster and are  missing in the
CF. To  quantify the  observed difference we  count the  stars falling
into the SB (with the additional conditions $(J-K)_0>0.95$ and $K_0\le
12.0$)  in the  two  CMDs ($N_{b}^{clus}$  and  $N_{b}^{CF}$, for  the
cluster field  and for the CF, respectively).   To take simultaneously
into  account  the  differences  in  sampled  area  and  any  possible
difference due  to galactic  population gradients between  the cluster
fields and  the CFs we normalize  to the number of  stars spanning the
same   magnitude   range  of   the   plotted   SB   and  with   colors
$0.55<(J-K)_0<0.90$               ($N_{norm}^{clus}$               and
$N_{norm}^{CF}$).  Note,  however,  that  all the  results  are
fully  confirmed if we  normalize by  the ratio  of the  sampled areas
instead. In particular, the normalization ratios by number and by area
are  virtually identical in  all of  the considered  cases.  Defined
$R=N_{norm}^{clus}/N_{norm}^{CF}$,    we   compute    the   normalized
difference:
\begin{equation}
\Delta N_{\star} = N_{b}^{clus} - R \times N_{b}^{CF} \, ,
\end{equation}
and its standard deviation, assuming Poisson errors in star counts:
\begin{equation}
\sigma(\Delta N_{\star}) = \sqrt{N_{b}^{clus} + R^2 \times N_{b}^{CF}} \, .
\end{equation}

In the  present case $\Delta N_{\star}  = 53.5 \pm  10.9$ is obtained,
that is,  a very significant  difference ($4.9 \sigma$).  The searched
RGB  feature is  indeed  present around  Pal~12,  it is  statistically
significant and it is found at the {\em right} magnitude level.

To  explore better the  nature of  the detected  signal we  repeat the
counts shifting  the SB in the  range $\Delta\mu_0 =  -0.76\pm 1.0$ at
0.04  mag  steps.  We  plot  the  resulting  $\Delta N_{\star}$  as  a
function of $\Delta\mu_0$ in the lower panel of Fig.~2. The normalized
difference reaches a  wide maximum in the range  $-1.1 \le \Delta\mu_0
\le -0.5$ and the characteristic $\Delta\mu_0$ of Pal~12 (indicated by
a filled triangle  in the top of the panel) falls  right in the middle
of  the  detected  plateau. The  plot  has  to  be considered  just  a
consistency  test  since  the  constraining  power on  distance  of  a
sparsely   populated    upper   RGB   sequence    is   obviously   not
strong. Moreover, the signal remains  significant over a wide range of
$\Delta N_{\star}$  because, after all,  the Sgr stars are  present in
the Pal~12 field  and absent in the  CF in this area of  the CMD. Thus
any  position of  the  SB that  includes  some of  the  Sgr RGB  stars
provides a sensible difference with respect to the CF.  The effects of
this syndrome  may be  particularly strong in  shifting the  SB toward
brighter magnitudes ($K_0$  \ltsima 10.5) since in this  region the CF
is virtually devoid of stars.

Notwithstanding the  above caveats the  detection of Sgr  Stream stars
around  Pal~12 is clear  and significant,  confirming the  findings by
\cite{dav12}. Given  it's position in the phase-space  with respect to
the orbit of the Sgr dSph (see BFI03) and its physical location within
the Sgr Stream,  we conclude that Pal~12 presently  belongs to the Sgr
Stream and that it was a  former member of the globular cluster system
of the Sgr dSph.

   \begin{figure*}
   \centering
   \ifthenelse{\CompactFigs=0}
   {\includegraphics[width=18cm]{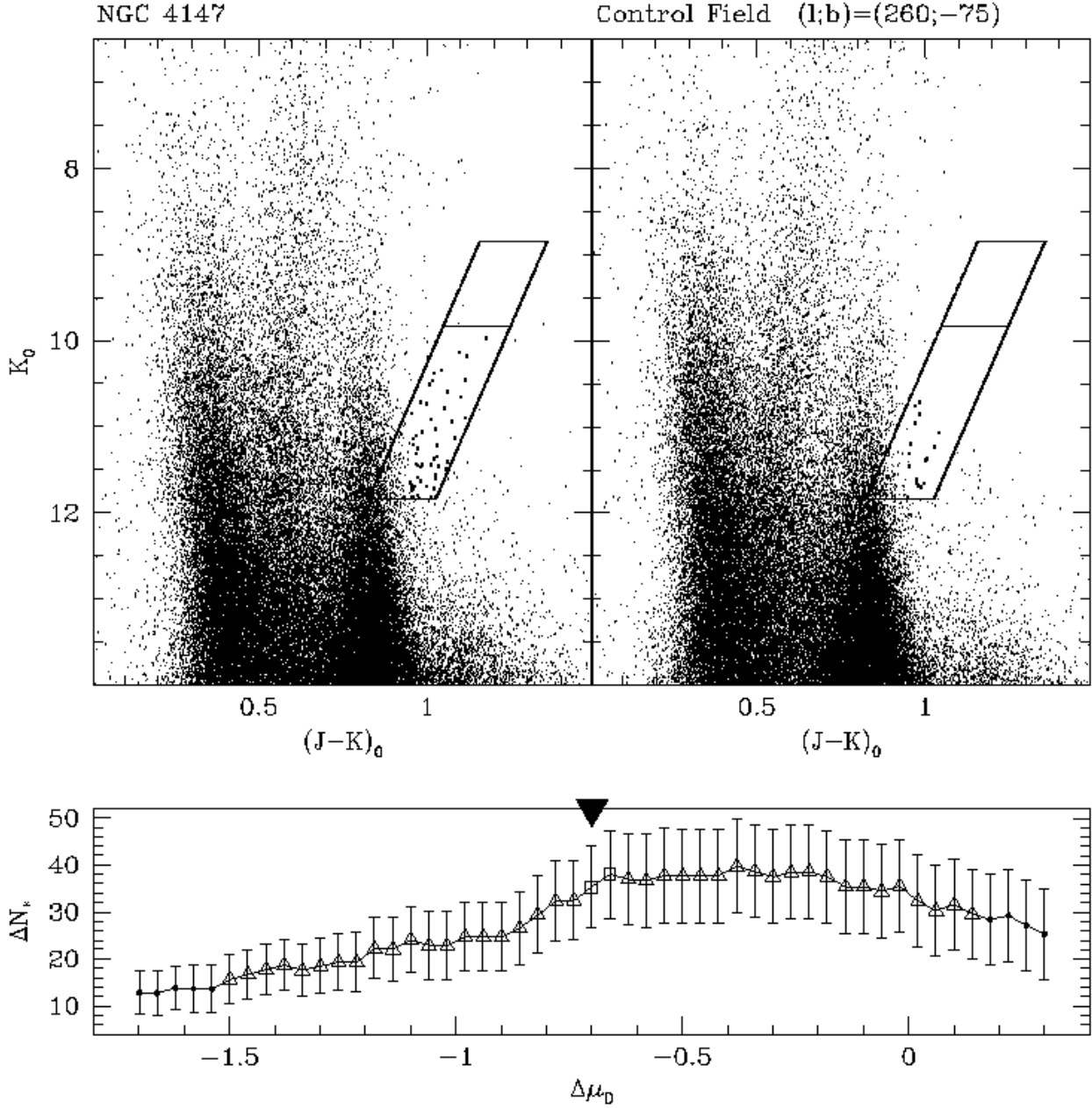}}
   {\includegraphics[width=18cm]{fig3.ps}}{}
   \caption{The same as Fig.~2 for the selected area around the position of
   NGC~4147 and of a Control Field symmetrical with respect to the Galactic
   equator. $\Delta \mu_0 = -0.70$, in this case.}
    \end{figure*}
%

Even an educated guess of the surface brightness ($\Sigma$) of the Sgr
Stream  at the  considered locations  would be  of great  interest for
future   modeling of  the dynamical  history of  the dwarf  galaxy. We
obtain such  an estimate  by counting  the number of  stars in  the SB
having  luminosities  within  1  magnitude  of  the  assumed  RGB  Tip
($N_{Tip}^{clus}$), dividing by the  sampled area and rescaling by the
central surface brightness of Sgr dSph to the ratio of surface density
of RGB Tip stars in the considered area and in the center of Sgr.  The
underlying assumption is that the  stellar population in the center of
the Sgr dSph and in the Sgr Stream are similar.  The adopted method is
widely  used  (see,   for  example,  \citealt{kleyna,mat91})  and  its
physical basis  is anchored on  the {\em Evolutionary  Flux Theorem}
(\citealt{rb86,rf88,alvio}).   In  the  central $0.5\times  0.5  ~{\rm
deg}^2$ of  Sgr dSph  we find $N_{Tip}^{clus}=50$  stars within  1 mag
from the  RGB Tip and we  assume that this corresponds  to the central
surface   brightness   of   $\Sigma_V(0)\sim   25.4  \pm   0.3   ~{\rm
mag/arcsec^2}$ reported by  \cite{M98}.  The possible contamination by
field stars is tentatively removed with the same technique used above,
i.e.         considering       $N_{\star}^{Tip}=N_{Tip}^{clus}-R\times
N_{Tip}^{CF}$ instead of  $N_{Tip}^{clus}$ itself.  Finally, we obtain
for the  average surface brightness  in the Sgr Stream  around Pal~12,
$\Sigma_V  = 33.9  \pm 0.5~{\rm  mag/arcsec}^2$.  The  uncertainty has
been   estimated    by   1000   Montecarlo    simulations   in   which
$\Sigma_V^{Sgr}(0)$,   $N_{Tip}^{clus}$   and   $N_{Tip}^{Sgr}$   were
simultaneously  extracted  from   gaussian  distributions  having  the
observed  values as mean  and the  reported $1-\sigma$  uncertainty as
$\sigma$.

\section{NGC 4147}

NGC~4147 $[(l;b)=(252.850\degr;77.189\degr)]$ lies  in a region of the
sky ($220\degr\simlt  l \simlt 320\degr$)  where the orbit of  the Sgr
dSph  runs   approximately  at  constant   galactic  latitude  ($b\sim
75\degr$). We  take advantage  from this occurrence  by taking  a wide
selected  area  comprised  within  $230\degr\le l  \le  290\degr$  and
$65\degr\le b \le 85\degr$.  The  effectively sampled field is 50\% of
the selected area.   As Control Field we choose the  region of the sky
symmetric with  respect to the galactic equator,  i.e.  $230\degr\le l
\le 290\degr$  and $-85\degr\le b  \le -65\degr$.  The  resulting CMDs
are shown  in Fig.~3, which is arranged  in the same way  and with the
same symbols as Fig.~2, above. The difference in true distance modulus
between  NGC~4147  and   the  core  of  the  Sgr   galaxy  is  $\Delta
\mu_0=\mu_0(N4147)-\mu_0(M54)=-0.70\pm  0.15$  and  the  SB  has  been
shifted accordingly in these panels with respect to Fig.~1.

Also  in  this  case  the   Sgr  RGB  feature  is  quite  evident  and
significant. The normalized difference is $\Delta N_{\star} = 35.4 \pm
8.7$, a  $4.0 \sigma$ signal. The  lower panel of Fig.~3  shows a less
pronounced maximum  with respect to  the Pal~12 case, probably  due to
the  higher degree  of contamination  of  the considered  CMDs in  the
region $(J-K)_0>0.95$.  The adopted $\Delta \mu_0$  for NGC~4147 falls
at the  ``bright'' edge  of the plateau.   However if we  consider the
normalized    difference    in    terms   of    standard    deviations
($N_{\star}/\sigma$),  the  maximum  signal  is  detected  at  $\Delta
\mu_0=-0.70 \pm 0.04$, i.e. at the distance of NGC~4147.

Hence, it  has to be concluded  that also this cluster  is immersed in
the  Sgr  Stream  and  that  it is  physically  associated  with  this
structure.

Using the technique illustrated above we obtained $\Sigma_V = 34.0 \pm
0.6 ~{\rm mag/arcsec^2}$ for the Sgr Stream surrounding NGC~4147.  The
comparison with  the analogous estimate obtained in  the region around
Pal~12  suggests that  the  Stream maintains  nearly constant  surface
brightness along its whole extension (see also \citealt{mom98}).

\section{Other cases}

In this  section we shortly comment  on the other  candidates by BFI03
considered by the present study.  In  all cases we cannot reach a firm
conclusion, either because of a  too small effectively sampled area or
because of problems with interstellar extinction.

NGC~5634 and Pal~5  are relatively nearby in the  sky and have similar
$\Delta \mu_0$, and thus  may be studied simultaneously.  Fig.~4 shows
that the  coverage of 2MASS  in this region  of sky is quite  poor and
unfavorably placed with respect to  the considered clusters and to the
Sgr  orbit. Hence,  the  absence  of a  significant  detection in  the
selected area  shown in Fig.~4  cannot be considered as  evidence that
the clusters are  not associated with the Sgr  Stream.  Hopefully, the
question  should  be  resolved  when  future  releases  of  the  2MASS
catalogue are available.

The same problem  plagues the application of the  test to NGC~5053. In
this  case only  a barely  significant detection  is obtained  but the
effective  sampling  is  clearly  insufficient  and  unfavorable  (see
Fig.~5).

 In Fig.~6 we report, for comparison, the distribution of the sampled 
fields in the selected area around NGC~4147.

   \begin{figure}
   \centering
   \ifthenelse{\CompactFigs=0}
   {\includegraphics[width=9cm]{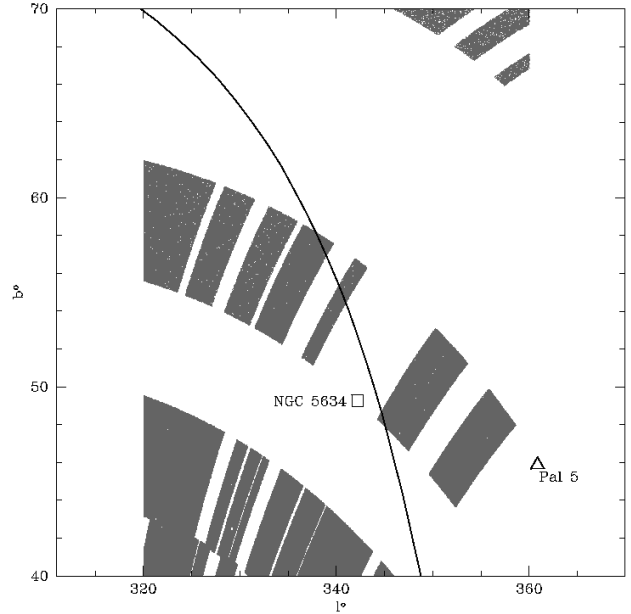}}
   {\includegraphics[width=9cm]{fig4.ps}}{}
   \caption{The fields sampled by 2MASS in the selected area around NGC~5634 and
   Pal~5 (in grey). The continuous line is the projection in the sky of the Sgr
   orbit by IL98.}
    \end{figure}
%

   \begin{figure}
   \centering
   \ifthenelse{\CompactFigs=0}
   {\includegraphics[width=9cm]{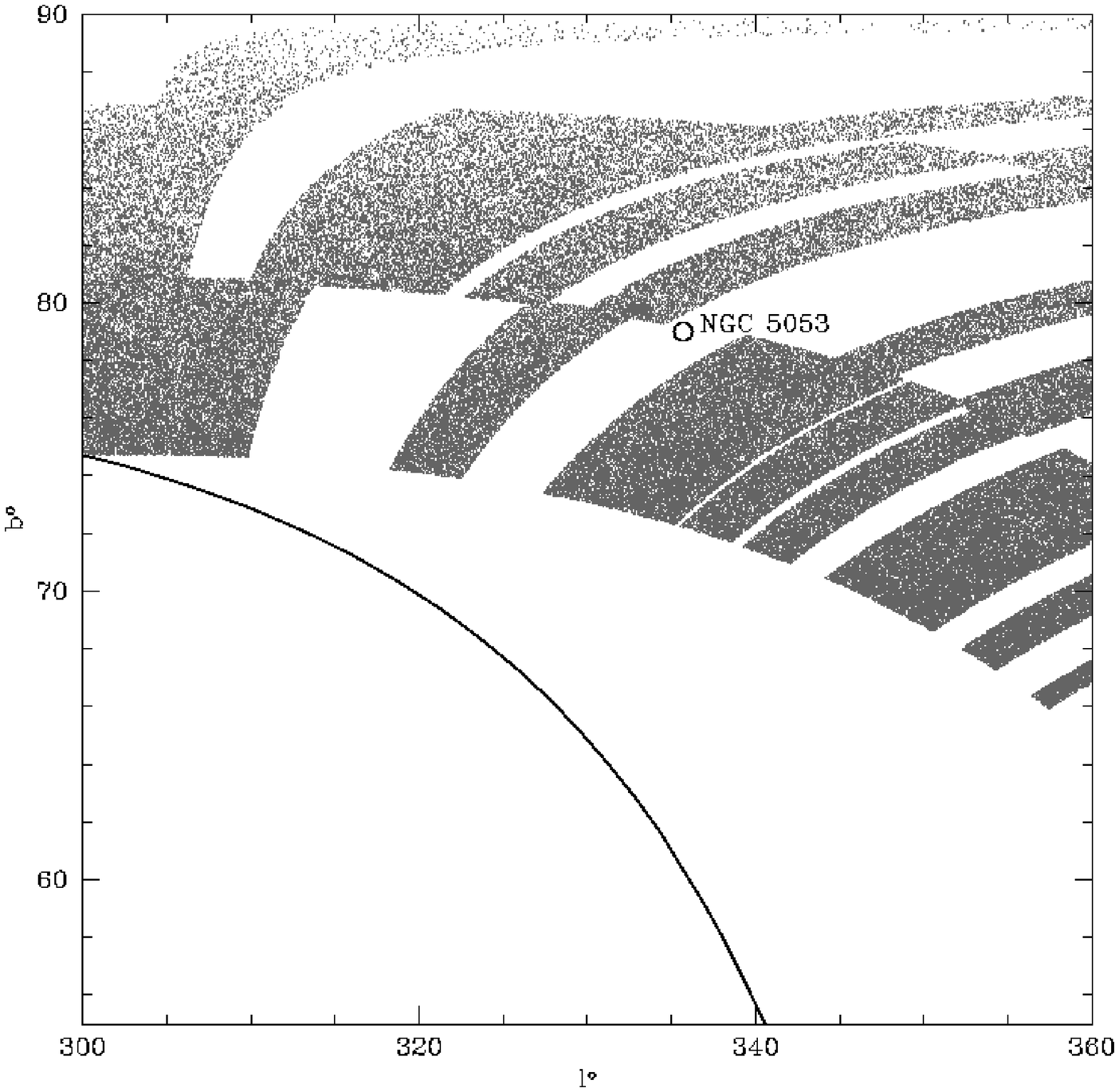}}
   {\includegraphics[width=9cm]{fig5.ps}}{}
   \caption{The fields sampled by 2MASS in the selected area around NGC~5053 
   (in grey). The continuous line is the projection in the sky of the Sgr
   orbit by IL98.}
    \end{figure}
%
   \begin{figure}
   \centering
   \ifthenelse{\CompactFigs=0}
   {\includegraphics[width=9cm]{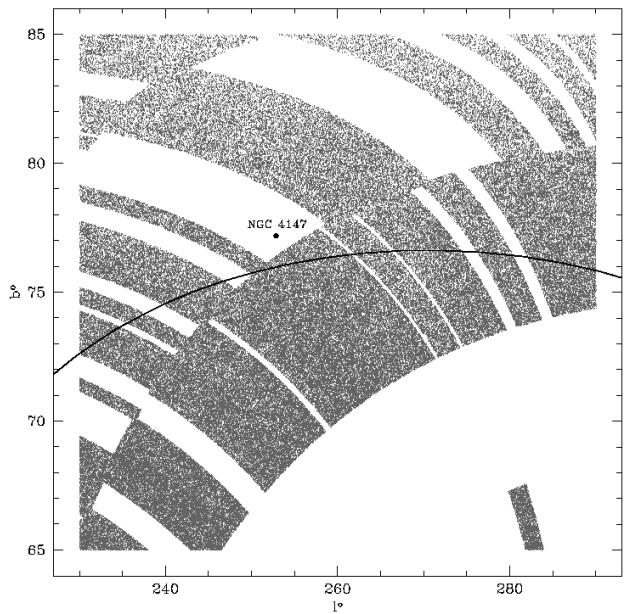}}
   {\includegraphics[width=9cm]{fig6.ps}}{}
   \caption{The fields sampled by 2MASS in the selected area around NGC~4147 
   (in grey). The continuous line is the projection in the sky of the Sgr
   orbit by IL98.}
    \end{figure}
%

\section{Summary and Conclusions}

 We have examined  the 2MASS $K_0$;$(J-K)_0$ CMDs of  wide areas of the
sky  around six  Outer  Halo  globular clusters  to  search for  stars
belonging to the Sgr Stream  and that are located approximately at the
same distance  as the considered  clusters.  The
considered clusters  were selected for their  phase-space proximity to
the orbit of  the Sgr galaxy, according to BFI03. 
The  dataset  was found  to be  suitable for  the
application of the test to the  cases of Pal~12 and NGC~4147.  In both
cases, an overdensity of tracer stars of similar heliocentric distance
as  the  clusters,  was  detected   at  the  $>4.5  \sigma$  level  of
significance with respect to Control Fields.  Hence we have shown that
both clusters are  physically immersed in the Sgr  Stream and are very
likely members  of this structure as  well as being  former members of
the Sgr dSph galaxy.  While  our result for Pal~12 confirms a previous
(independent) detection  by \cite{dav12}, this is the  first time that
Sgr Stream stars are found around NGC~4147.  
A thorough  analysis of the  structure and the stellar  populations of
the  Sgr Stream  will be  presented when  the full-sky  2MASS catalogs
become available.

\begin{acknowledgements}
M.B. and F.R.F. acknowledge the  financial support to this research by
the italian {Ministero  dell'Universit\'a e della Ricerca Scientifica}
(MURST) through the grant p.  2001028879, assigned to the project {\em
Origin and Evolution of Stellar Populations in the Galactic Spheroid}.
We are grateful to V. Marsakov for calling our attention to the new distance
modulus of Ter~3, and to the Referee (M. Mateo) for several useful comments.
This research has made use of NASA's Astrophysics Data System Abstract
Service. M.B. dedicates this work to the memory of his father Giocondo.
\end{acknowledgements}

{}


\begin{thebibliography}{}

\bibitem[Alard(2001)]{alard} Alard, C., 2001, A\&A, 377, 389
\bibitem[Bellazzini, Ferraro \& Ibata(2003)]{pap1} Bellazzini, M., Ferraro,
         F.R. \& Ibata, R., 2003, AJ, 125, 188 [BFI03]
\bibitem[Cole(2001)]{cole} Cole, A.A., 2001, ApJ, 559, L17	 
\bibitem[Dohm-Palmer et al.(2001)]{dohm} Dohm-Palmer, R.C., Helmi, A., Morrison,
         H., Mateo, M., Olszewski, E.W., Harding, P., Freeman, K.C., Norris, J.
	 \& Shectman, S.A., 2001, ApJ, 555, L37
\bibitem[Dinescu et al.(2000)]{pal12} Dinescu, D.I., Majewski, S.R., 
         Girard, T.M. \& Cudworth, K.M., 2000, AJ, 120, 1892
\bibitem[Fahlman et al.(1996)]{sgrm55} Fahlman, G.G., Mandushev, G., Richer,
         H.B., Thompson, I.B. \& Sivaranakrishnan, A., 1996, ApJ, 459, L65	 
\bibitem[Harris(1996)]{harris} Harris, W.E. 1996, AJ, 112, 1487
\bibitem[Helmi et al.(1999)]{ami1} Helmi, A., White, S., de Zeeuw, P.T. \& Zhao,
         H., 1999, Nature, 402, 53
\bibitem[Kundu et al.(2002)]{kundu} Kundu, A., et al., 2002, ApJ, L125	 
\bibitem[Ibata et al.(1994)]{s1} Ibata, R.A., Irwin, M.J. \& Gilmore, G., 1994,
         Nature, 370, 194
\bibitem[Ibata et al.(1997)]{s2} Ibata, R.A., Wyse, R.F.G. \& Gilmore, G., 
         Irwin, M.J., \& Suntzeff, N.B., 1997, AJ, 113, 634
\bibitem[Ibata \& Lewis(1998)]{orb} Ibata, R.A. \& Lewis, G.F., 1998, ApJ,
         500, 575 (IL98)	 
\bibitem[Ibata et al.(2001a)]{carb} Ibata, R.A., Lewis, G.F., Irwin, M., 
         Totten, E. \& Quinn, T., 2001a, ApJ, 551, 294
\bibitem[Ibata et al.(2001b)]{orb-sds} Ibata, R.A., Irwin, M., Lewis, G.F. \&	\&
         Stolte, A., 2001b, ApJ, 547, L136 
\bibitem[Ibata et al.(2002)]{orb-2mass}	Ibata, R.A., Lewis, G.F., Irwin, M. \&
         Cambr\'esy, L., 2002, MNRAS, 332, 921  
\bibitem[Ivezic et al.(2000)]{ivez} Ivezic, Z., et al., 2000, AJ, 120, 963 
\bibitem[Johnston et al.(1999)]{katy2} Johnston, K.V., Zhao, H., Spergel, D.N. 
         \& Hernquist, L.,1999, ApJ, 512, L109
\bibitem[Kleyna et al.(1998)]{kleyna} Kleyna, J.T., Geller, M.J., Kenyon, S.J.,
         Kurtz, M.J. \& Thorstensen, J.R., 1998, AJ, 115, 2368	 
\bibitem[Majewski et al.(1999)]{maj99} Majewski, S.R., Siegel, M.H., Kunkel, W.E.,
         Reid, I.N., Johnston, K.V., Thompson, I.B., Landolt, A.U. \& Palma,
	 C., 1999, AJ, 118, 1709	 
\bibitem[Mart\'inez-Delgado et al.(2001)]{david} Mart\'inez-Delgado, D., 
         Aparicio, A., G\'omez-Flechoso, M.A. \& Carrera, R., 2001, ApJ, 549,
	 L199
\bibitem[Mart\'inez-Delgado et al.(2002)]{dav12} Mart\'inez-Delgado, D., Zinn,
         R., Carrera, R. \& Gallart, C., 2002, ApJ, 537, L19	 
\bibitem[Mateo(1998)]{M98} Mateo, M., 1998, ARA\&A, 36, 435
\bibitem[Mateo et al.(1991)]{mat91} Mateo, M., Nemec, J., Irwin, M. \& McMahon,
         R., AJ, 101, 892	 	 
\bibitem[Mateo, Olszewski \& Morrison(1998)]{mom98}Mateo, M., Olszewski, E.W. 
         \& Morrison, H.L., 1998, ApJ, 508, L55
\bibitem[Monaco et al.(2002)]{lorenzo} Monaco, L., Ferraro, F.R., Bellazzini,
         M. \& Pancino, E., 2002, ApJ, 578, L47	 
\bibitem[Newberg et al.(2002)]{sdss} Newberg, H.J., et al., 2002, ApJ, 569, 245
\bibitem[Renzini(1998)]{alvio} Renzini, A., 1998, AJ, 115, 2465
\bibitem[Renzini \& Buzzoni(1986)]{rb86} Renzini, A., \& Buzzoni, A., 1986, in
         Spectral Evolution of Galaxies, eds. C. Chiosi and A. Renzini,
	 (Drodrecht: Reidel), 135
\bibitem[Renzini \& Fusi Pecci(1988)]{rf88} Renzini, A. \& Fusi Pecci, F., 1988,
         ARA\&A, 26, 199	 
\bibitem[Savage \& Mathis(1979)]{sm79}Savage, B.D. \& Mathis, J.S., 1979,
         ARA\&A, 17,73 
\bibitem[Schlegel, Finkbeiner \& Davis(1998)]{iras}Schlegel, D., Finkbeiner, D.
         \& Davis, M., 1998, ApJ, 500, 525
\bibitem[Vivas et al.(2001)]{vivas} Vivas, A.K., et al., 2001, ApJ, 554, L33	 	 
\bibitem[Yanny et al.(2000)]{yanni} Yanny, B., et al., 2000, ApJ, 540, 825
\bibitem[Yanny et al.(2003)]{yanni2} Yanny, B., et al., 2003, ApJ, in press
        (astro-ph/0301029)
\end{thebibliography}
\end{document}